\documentclass[10pt, sigplan]{acmart}
\usepackage[utf8]{inputenc}
\usepackage[T1]{fontenc}

\newcommand{\packageGraphicx}{\usepackage{graphicx}}
\newcommand{\packageHyperref}{\usepackage{hyperref}}
\newcommand{\renewrmdefault}{\renewcommand{\rmdefault}{ptm}}
\newcommand{\packageRelsize}{\usepackage{relsize}}
\newcommand{\packageAmsmath}{\usepackage{amsmath}}
\newcommand{\packageMathabx}{\usepackage{mathabx}}
\newcommand{\packageWasysym}{
  \let\leftmoon\relax \let\rightmoon\relax \let\fullmoon\relax \let\newmoon\relax \let\diameter\relax
  \usepackage[nointegrals]{wasysym}}
\newcommand{\packageTxfonts}{
  \let\widering\relax
  \let\oldwidebar\widebar
  \let\widebar\relax
  \usepackage{newtxmath}
  \ifx\widebar\relax
    \let\widebar\oldwidebar
  \fi
}
\newcommand{\packageTextcomp}{\usepackage{textcomp}}
\newcommand{\packageFramed}{\usepackage{framed}}
\newcommand{\packageHyphenat}{\usepackage[htt]{hyphenat}}
\newcommand{\packageColor}{\usepackage[usenames,dvipsnames]{color}}
\newcommand{\doHypersetup}{\hypersetup{bookmarks=true,bookmarksopen=true,bookmarksnumbered=true}}
\newcommand{\packageTocstyle}{}
\newcommand{\packageCJK}{\IfFileExists{CJK.sty}{\usepackage{CJK}}{}}
\renewcommand\packageColor\relax
\renewcommand\packageTocstyle\relax
\renewcommand\packageMathabx{\ifx\bigtimes\undefined \usepackage{mathabx} \else \relax \fi}
\renewcommand\packageTxfonts\relax

\renewcommand{\renewrmdefault}{}

\packageGraphicx
\packageHyperref
\renewrmdefault
\packageRelsize
\packageAmsmath
\packageMathabx
\packageWasysym
\packageTxfonts
\packageTextcomp
\packageFramed
\packageHyphenat
\packageColor
\doHypersetup
\packageTocstyle
\packageCJK

\newcommand{\sectionNewpage}{}

\newcommand{\preDoc}{}
\newcommand{\postDoc}{}

\newcommand{\BookRef}[2]{\emph{#2}}
\newcommand{\ChapRef}[2]{\SecRef{#1}{#2}}
\newcommand{\SecRef}[2]{section~#1}
\newcommand{\PartRef}[2]{part~#1}
\newcommand{\BookRefUC}[2]{\BookRef{#1}{#2}}
\newcommand{\ChapRefUC}[2]{\SecRefUC{#1}{#2}}
\newcommand{\SecRefUC}[2]{Section~#1}
\newcommand{\PartRefUC}[2]{Part~#1}

\newcommand{\BookRefLocal}[3]{\hyperref[#1]{\BookRef{#2}{#3}}}
\newcommand{\ChapRefLocal}[3]{\hyperref[#1]{\ChapRef{#2}{#3}}}
\newcommand{\SecRefLocal}[3]{\hyperref[#1]{\SecRef{#2}{#3}}}
\newcommand{\PartRefLocal}[3]{\hyperref[#1]{\PartRef{#2}{#3}}}
\newcommand{\BookRefLocalUC}[3]{\hyperref[#1]{\BookRefUC{#2}{#3}}}
\newcommand{\ChapRefLocalUC}[3]{\hyperref[#1]{\ChapRefUC{#2}{#3}}}
\newcommand{\SecRefLocalUC}[3]{\hyperref[#1]{\SecRefUC{#2}{#3}}}
\newcommand{\PartRefLocalUC}[3]{\hyperref[#1]{\PartRefUC{#2}{#3}}}

\newcommand{\BookRefUN}[1]{\BookRef{}{#1}}

\newcommand{\SecRefUN}[1]{``#1''}

\newcommand{\BookRefLocalUN}[2]{\hyperref[#1]{\BookRefUN{#2}}}

\newcommand{\SecRefLocalUN}[2]{\hyperref[#1]{\SecRefUN{#2}}}

\newcommand{\SectionNumberLink}[2]{\hyperref[#1]{#2}}

\newcommand{\Scribtexttt}[1]{{\texttt{#1}}}

\newcommand{\planetName}[1]{PLane\hspace{-0.1ex}T}

\makeatletter
\def\empty@finalstrut#1{%
  \unskip\ifhmode\nobreak\fi\vrule\@width\z@\@height\z@\@depth\z@}
\def\no@strut{\global\setbox\@arstrutbox\hbox{%
    \vrule \@height\z@
           \@depth\z@
           \@width\z@}%
    \gdef\@endpbox{\empty@finalstrut\@arstrutbox\par\egroup\hfil}%
}%
\def\yes@strut{\global\setbox\@arstrutbox\hbox{%
    \vrule \@height\arraystretch \ht\strutbox
           \@depth\arraystretch \dp\strutbox
           \@width\z@}%
    \gdef\@endpbox{\@finalstrut\@arstrutbox\par\egroup\hfil}%
}%
\def\@mkpream#1{\@firstamptrue\@lastchclass6
  \let\@preamble\@empty\def\empty@preamble{\add@ins}%
  \let\protect\@unexpandable@protect
  \let\@sharp\relax\let\add@ins\relax
  \let\@startpbox\relax\let\@endpbox\relax
  \@expast{#1}%
  \expandafter\@tfor \expandafter
    \@nextchar \expandafter:\expandafter=\reserved@a\do
       {\@testpach\@nextchar
    \ifcase \@chclass \@classz \or \@classi \or \@classii \or \@classiii
      \or \@classiv \or\@classv \fi\@lastchclass\@chclass}%
  \ifcase \@lastchclass \@acol
      \or \or \@preamerr \@ne\or \@preamerr \tw@\or \or \@acol \fi}
\def\@addamp{%
  \if@firstamp
    \@firstampfalse
    \edef\empty@preamble{\add@ins}%
  \else
    \edef\@preamble{\@preamble &}%
    \edef\empty@preamble{\expandafter\noexpand\empty@preamble &\add@ins}%
  \fi}
\newif\iftw@hlines \tw@hlinesfalse
\def\@xhline{\ifx\reserved@a\hline
               \tw@hlinestrue
             \else\ifx\reserved@a\Hline
               \tw@hlinestrue
             \else
               \tw@hlinesfalse
             \fi\fi
      \iftw@hlines
        \aftergroup\do@after
      \fi
      \ifnum0=`{\fi}%
}
\def\do@after{\emptyrow[\the\doublerulesep]}
\def\emptyrow{\noalign\bgroup\@ifnextchar[\@emptyrow{\@emptyrow[\z@]}}
\def\@emptyrow[#1]{\no@strut\gdef\add@ins{\vrule \@height\z@ \@depth#1 \@width\z@}\egroup%
\empty@preamble\\
\noalign{\yes@strut\gdef\add@ins{\vrule \@height\z@ \@depth\z@ \@width\z@}}%
}
\def\tabrow#1{\noalign\bgroup\@ifnextchar[{\@tabrow{#1}}{\@tabrow{#1}[]}}
\def\@tabrow#1[#2]{\no@strut\egroup#1\ifx.#2.\\\else\\[#2]\fi\noalign{\yes@strut}}
\def\endpltstabular{\crcr\egroup\egroup \egroup}
\expandafter \let \csname endpltstabular*\endcsname = \endpltstabular
\def\pltstabular{\let\@halignto\@empty\@pltstabular}
\@namedef{pltstabular*}#1{\def\@halignto{to#1}\@pltstabular}
\def\@pltstabular{\leavevmode \bgroup \let\@acol\@tabacol
   \let\@classz\@tabclassz
   \let\@classiv\@tabclassiv \let\\\@tabularcr\@stabarray}
\def\@stabarray{\m@th\@ifnextchar[\@sarray{\@sarray[c]}}
\def\@sarray[#1]#2{%
  \bgroup
  \setbox\@arstrutbox\hbox{%
    \vrule \@height\arraystretch\ht\strutbox
           \@depth\arraystretch \dp\strutbox
           \@width\z@}%
  \@mkpream{#2}%
  \edef\@preamble{%
    \ialign \noexpand\@halignto
      \bgroup \@arstrut \@preamble \tabskip\z@skip \cr}%
  \let\@startpbox\@@startpbox \let\@endpbox\@@endpbox
  \let\tabularnewline\\%
    \let\@sharp##%
    \set@typeset@protect
    \lineskip\z@skip\baselineskip\z@skip
    \@preamble}

\makeatother

\newenvironment{bigtabular}{\begin{pltstabular}}{\end{pltstabular}}

\newlength{\stabLeft}
\newcommand{\bigtableleftpad}{\hspace{\stabLeft}}

\newenvironment{SingleColumn}{\begin{list}{}{\topsep=0pt\partopsep=0pt%
\listparindent=0pt\itemindent=0pt\labelwidth=0pt\leftmargin=0pt\rightmargin=0pt%
\itemsep=0pt\parsep=0pt}\item}{\end{list}}

\newcommand{\SCodePreSkip}{\vskip\abovedisplayskip}
\newcommand{\SCodePostSkip}{\vskip\belowdisplayskip}

\newcommand{\SVInsetPreSkip}{\vskip\abovedisplayskip}
\newcommand{\SVInsetPostSkip}{\vskip\belowdisplayskip}

\newenvironment{SCentered}{\begin{trivlist}\item \centering}{\end{trivlist}}

\newcommand{\titleAndVersionAndAuthors}[3]{\title{#1\\{\normalsize \SVersionBefore{}#2}}\author{#3}\maketitle}

\newcommand{\titleAndEmptyVersionAndAuthors}[3]{\title{#1}\author{#3}\maketitle}

\newcommand{\SAuthor}[1]{#1}
\newcommand{\SAuthorSep}[1]{\qquad}
\newcommand{\SVersionBefore}[1]{Version }

\newcommand{\SNumberOfAuthors}[1]{}

\let\SOriginalthesubsection\thesubsection
\let\SOriginalthesubsubsection\thesubsubsection

\newcommand{\SNextTitlePlain}[1]{\def\theNextTitlePlain{#1}}
\newcommand{\STexOrPDFTitle}[1]{\texorpdfstring{#1}{\theNextTitlePlain}}

\newcommand{\Ssection}[2]{\section[\STexOrPDFTitle{#1}]{#2}\let\thesubsection\SOriginalthesubsection}
\newcommand{\Ssubsection}[2]{\subsection[\STexOrPDFTitle{#1}]{#2}\let\thesubsubsection\SOriginalthesubsubsection}

\newcommand{\Ssectionstar}[1]{\section*{#1}\renewcommand*\thesubsection{\arabic{subsection}}\setcounter{subsection}{0}}

\newcommand{\Ssectionstarx}[2]{\Ssectionstar{#2}\phantomsection\addcontentsline{toc}{section}{\STexOrPDFTitle{#1}}}

\newcounter{GrouperTemp}

\newenvironment{SVerbatim}{}{}

\newcommand{\Snolinkurl}[1]{\nolinkurl{#1}}

\newcommand{\SAuthorinfo}[4]{#1}
\newcommand{\SAuthorPlace}[1]{#1}
\newcommand{\SAuthorEmail}[1]{#1}

\newcommand{\SConferenceInfo}[2]{}
\newcommand{\SCopyrightYear}[1]{}
\newcommand{\SCopyrightData}[1]{}
\newcommand{\Sdoi}[1]{}

\newcommand{\SCategory}[3]{}
\newcommand{\SCategoryPlus}[4]{}
\newcommand{\STerms}[1]{}
\newcommand{\SKeywords}[1]{}

\newcommand{\AutobibLink}[1]{\color{ACMPurple}{#1}}

\newcommand{\Autobibtarget}[1]{\phantomsection#1}

\usepackage{calc}
\newlength{\ABcollength}
\newcommand{\Autocolbibnumber}[1]{\parbox[t]{5ex}{\hfill#1~~\vspace{1.0ex}}}
\newcommand{\Autocolbibentry}[1]{\setlength{\ABcollength}{\linewidth-5ex}\parbox[t]{\ABcollength}{#1\vspace{1.0ex}}}

\newcommand{\Autobibref}[1]{#1}

\providecommand{\AutobibLink}[1]{#1}

\newcommand{\NoteBox}[1]{\footnote{#1}}
\newcommand{\NoteContent}[1]{#1}

\newcommand{\FootnoteRef}[1]{}

\newcommand{\FootnoteTarget}[1]{}

\newcommand{\FootnoteBlockContent}[1]{}
\usepackage{ccaption}

\makeatletter
\@ifundefined{belowcaptionskip}{}{}
\makeatother

\newcommand{\Legend}[1]{~

                        \hrule width \hsize height .33pt
                        \vspace{4pt}
                        \legend{#1}}

\newcommand{\FigureTarget}[2]{#1}

\newlength{\FigOrigskip}
\FigOrigskip=\parskip

\newcommand{\FigureSetRef}{\refstepcounter{figure}}

\newenvironment{Figure}{\begin{figure}\FigureSetRef}{\end{figure}}
\newenvironment{FigureMulti}{\begin{figure*}[t!p]\FigureSetRef}{\end{figure*}}

\newenvironment{Centerfigure}{\begin{Xfigure}\centering\item}{\end{Xfigure}}

\newenvironment{Xfigure}{\begin{list}{}{\leftmargin=0pt\topsep=0pt\parsep=\FigOrigskip\partopsep=0pt}}{\end{list}}

\newenvironment{FigureInside}{}{}

\newcommand{\Centertext}[1]{\begin{center}#1\end{center}}

\renewcommand{\titleAndVersionAndAuthors}[3]{\title{#1}#3\maketitle}
\renewcommand{\titleAndEmptyVersionAndAuthors}[3]{\titleAndVersionAndAuthors{#1}{#2}{#3}}

\def\SAuthor#1{\SAutoAuthor#1\SAutoAuthorDone{#1}}
\def\SAutoAuthorDone#1{}
\def\SAutoAuthor{\futurelet\next\SAutoAuthorX}
\def\SAutoAuthorX{\ifx\next\SAuthorinfo \let\Snext\relax \else \let\Snext\SToAuthorDone \fi \Snext}
\def\SToAuthorDone{\futurelet\next\SToAuthorDoneX}
\def\SToAuthorDoneX#1{\ifx\next\SAutoAuthorDone \let\Snext\SAddAuthorInfo \else \let\Snext\SToAuthorDone \fi \Snext}
\newcommand{\SAddAuthorInfo}[1]{\SAuthorinfo{#1}{}{}}

\renewcommand{\SAuthorinfo}[4]{\author{#1}{#2}{#3}{#4}}
\renewcommand{\SAuthorSep}[1]{}

\renewcommand{\SAuthorPlace}[1]{\affiliation{#1}}
\renewcommand{\SAuthorEmail}[1]{\email{#1}}

\renewcommand{\SConferenceInfo}[2]{\conferenceinfo{#1}{#2}}
\renewcommand{\SCopyrightYear}[1]{\copyrightyear{#1}}
\renewcommand{\SCopyrightData}[1]{\copyrightdata{#1}}

\renewcommand{\SCategory}[3]{\category{#1}{#2}{#3}}
\renewcommand{\SCategoryPlus}[4]{\category{#1}{#2}{#3}[#4]}
\renewcommand{\STerms}[1]{\terms{#1}}
\renewcommand{\SKeywords}[1]{\keywords{#1}}

\newcommand{\SccsdescNumber}[2]{\ccsdesc[#1]{#2}}
\begin{document}
\preDoc

\settopmatter{printccs=true, printacmref=false, printfolios=false}

\setcopyright{none}

\copyrightyear{}


\startPage{42}

\acmConference{Off the Beaten Track}{2018}{Los Angeles}

\acmYear{}

\SccsdescNumber{300}{Software and its engineering~Reflective middleware}

\acmPrice{}

\acmDOI{}

\acmISBN{}

\SccsdescNumber{500}{Theory of computation~Operational semantics}

\SccsdescNumber{500}{Theory of computation~Categorical semantics}

\SccsdescNumber{100}{Theory of computation~Type theory}

\SccsdescNumber{500}{Software and its engineering~Reflective middleware}

\SccsdescNumber{500}{Software and its engineering~Runtime environments}

\SccsdescNumber{300}{Software and its engineering~Just-in-time compilers}


\begin{abstract}Software exists at multiple levels of abstraction,
where each more concrete level is an implementation of the more abstract level above,
in a semantic tower of compilers and/or interpreters.
First{-}class implementations are a reflection protocol to navigate this tower \emph{at runtime}:
they enable changing the underlying implementation of a computation \emph{while it is running}.
Key is a generalized notion of \emph{safe points}
that enable observing a computation at a higher{-}level than that at which it runs,
and therefore to climb up the semantic tower,
when at runtime most existing systems only ever allow but to go further down.
The protocol was obtained by extracting the computational content of a formal specification
for implementations and some of their properties.
This approach reconciles two heretofore mutually exclusive fields:
Semantics and Runtime Reflection.\end{abstract}\titleAndEmptyVersionAndAuthors{Climbing Up the Semantic Tower {---} at Runtime}{}{\SNumberOfAuthors{1}\SAuthor{\SAuthorinfo{Fran\c{c}ois{-}Ren\'{e} Rideau}{}{\SAuthorPlace{\institution{TUNES}\city{}\country{}}}{\SAuthorEmail{fare@tunes.org}}}}
\label{t:x28part_x22Climbingx5fUpx5fthex5fSemanticx5fTowerx5fx2dx2dx2dx5fatx5fRuntimex22x29}

\noindent 

\noindent

\noindent 

\noindent 

\noindent 

\noindent 

\noindent 

\noindent 

\noindent 

\noindent

\noindent 

\noindent 

\noindent 

\noindent 

\noindent

\SNextTitlePlain{Introduction}\sectionNewpage

\Ssection{Introduction}{Introduction}\label{t:x28part_x22Introductionx22x29}

Semantics predicts properties of computations without running them.
Runtime Reflection allows unpredictable modifications to running computations.
The two seem opposite, and those who practice one tend to ignore or prohibit the other.
This work\Autobibref{~[\hyperref[t:x28autobib_x22Franxe7oisx2dRenxe9_RideauReconciling_Semantics_and_Reflection2018x22x29]{\AutobibLink{6}}]} reconciles them:
semantics can specify \emph{what} computations do,
reflection can control \emph{how} they do it.

\SNextTitlePlain{Formalizing Implementations}\sectionNewpage

\Ssection{Formalizing Implementations}{Formalizing Implementations}\label{t:x28part_x22Formalizingx5fImplementationsx22x29}

An elementary use of Category Theory can unify
Operational Semantics and other common model of computations:
potential states of a computation and labelled transitions between them
are the nodes ({``}objects{''}) and arrows ({``}morphisms{''}) of a category.
The implementation of an abstract computation $A$ with a concrete one $C$
is then a {``}partial functor{''} from $C$ to $A$, i.e.
given a subset $O$ of {``}observable{''} safe points in $C$,
a span of an interpretation functor from $O$ to $A$
and the full embedding of $O$ in $C$.
\emph{Partiality} is essential: concepts atomic in an abstract calculus
usually are not atomic in a more concrete calculus;
concrete computations thus include many intermediate steps
not immediately meaningful in the abstract.\NoteBox{\NoteContent{For instance, languages in the ALGOL tradition have no notion of explicit data registers or stacks,
yet are typically implemented using lower{-}level machines (virtual or {``}real{''}) that do;
meanwhile their high{-}level {``}primitives{''} each require many low{-}level instructions to implement.}}

\begin{Figure}\begin{Centerfigure}\begin{FigureInside}\begin{bigtabular}{@{\bigtableleftpad}c@{}c@{}c@{}c@{}c@{}c@{}c@{}}
\hbox{\includegraphics[scale=0.55]{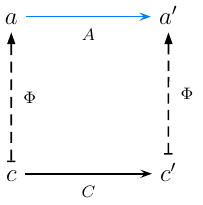}} &
\hbox{~~~~~\hfill~~~~~~} &
\hbox{\includegraphics[scale=0.55]{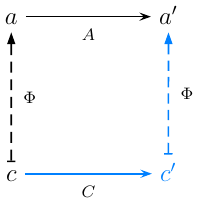}} &
\hbox{~~~~~\hfill~~~~~~} &
\hbox{\includegraphics[scale=0.55]{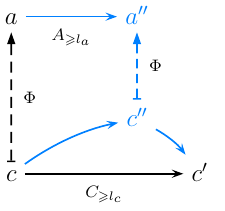}} &
\hbox{~~~~~\hfill~~~~~~} &
\hbox{\includegraphics[scale=0.55]{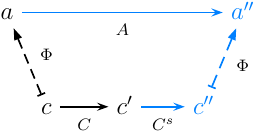}} \\
\begin{minipage}[c]{0.14285714285714285\linewidth}
\begin{SCentered}Sound\end{SCentered} \end{minipage}
 &
\hbox{~~~~~\hfill~~~~~~} &
\begin{minipage}[c]{0.14285714285714285\linewidth}
\begin{SCentered}Complete\end{SCentered} \end{minipage}
 &
\hbox{~~~~~\hfill~~~~~~} &
\begin{minipage}[c]{0.14285714285714285\linewidth}
\begin{SCentered}Live\end{SCentered} \end{minipage}
 &
\hbox{~~~~~\hfill~~~~~~} &
\begin{minipage}[c]{0.14285714285714285\linewidth}
\begin{SCentered}Observable\end{SCentered} \end{minipage}
\end{bigtabular}\end{FigureInside}\end{Centerfigure}

\Centertext{\Legend{\FigureTarget{\label{t:x28counter_x28x22figurex22_x22figx2dpropertiesx22x29x29}\textsf{Fig.}~\textsf{1}. }{t:x28counter_x28x22figurex22_x22figx2dpropertiesx22x29x29}\textsf{Some properties for implementations to have or not}}}\end{Figure}

The mandatory \emph{soundness} criterion is, remarkably, the same as functoriality.
Many other interesting properties may or may not hold for a given implementation:
variants of \emph{completeness} guarantee that
abstract nodes or arrows are not left unimplemented in the concrete
(e.g. can express the notion of simulation\Autobibref{~[\hyperref[t:x28autobib_x22Robin_MilnerAn_Algebraic_Definition_of_Simulation_between_Programs1971x22x29]{\AutobibLink{4}}]});
variants of \emph{liveness} guarantee that progress in the abstract is made
given enough progress in the concrete
(e.g. can express {``}real time{''} behavior);
and variants of \emph{observability} guarantee that an observable abstract state
can be recovered given any intermediate state at which the concrete computation is interrupted.
These properties can be visualized using bicolor diagrams such as in figure 1.\NoteBox{\NoteContent{In these diagrams, computation is from left to right; abstract is above and concrete below;
property premises are in black and conclusions in blue; and all diagrams commute.
While an implementation is notionally from abstract to concrete, the opposite arrows
of Abstract Interpretation are drawn, because functoriality goes from concrete to abstract,
which is what matters when diagrams commute;
for more details on the diagrams see\Autobibref{~[\hyperref[t:x28autobib_x22Franxe7oisx2dRenxe9_RideauReconciling_Semantics_and_Reflection2018x22x29]{\AutobibLink{6}}]}.}}

\SNextTitlePlain{Extracting a Runtime Protocol}\sectionNewpage

\Ssection{Extracting a Runtime Protocol}{Extracting a Runtime Protocol}\label{t:x28part_x22Extractingx5fax5fRuntimex5fProtocolx22x29}

The above properties can be formalized using dependent types;
their constructive proofs will then have a \emph{computational content}
as per the Curry{-}Howard Correspondence\Autobibref{~[\hyperref[t:x28autobib_x22William_Ax2e_HowardThe_formulaex2dasx2dtypes_notion_of_construction1980x22x29]{\AutobibLink{3}}]}.
\emph{Observability} could thus be formalized in Agda\Autobibref{~[\hyperref[t:x28autobib_x22Ulf_NorellDependently_typed_programming_in_Agda2008x22x29]{\AutobibLink{5}}]}
as the type of the following function \Scribtexttt{observe} where:
\Scribtexttt{{\hbox{\texttt{.}}}o} and \Scribtexttt{{\hbox{\texttt{.}}}$\Rightarrow$} denote node{-}wise and arrow{-}wise components;
\Scribtexttt{$\Phi$} is the interpretation functor opposite the implementation of \Scribtexttt{A} with \Scribtexttt{C};
\Scribtexttt{a} is the starting abstract state concretely implemented by \Scribtexttt{c} (implicit inputs);
\Scribtexttt{c}\Scribtexttt{{'}}\Scribtexttt{} is the concrete state in which \Scribtexttt{C} was interrupted after effects \Scribtexttt{f} (explicit input);
\Scribtexttt{c}\Scribtexttt{{''}}\Scribtexttt{} is the observable safe point that is being recovered after effects \Scribtexttt{g} (explicit output);
\Scribtexttt{safe{\hbox{\texttt{.}}}$\Rightarrow$} guarantees that \Scribtexttt{g} cannot take too much resources or do blocking I/O
or require user intervention;
and \Scribtexttt{a}\Scribtexttt{{''}}\Scribtexttt{}, \Scribtexttt{h} and the last property ensure the diagram commutes (implicit outputs).

\noindent \begin{SVerbatim}\begin{SingleColumn}\Scribtexttt{observe {\hbox{\texttt{:}}} $\forall$ {\char`\{}a {\hbox{\texttt{:}}} A{\hbox{\texttt{.}}}o{\char`\}} {\char`\{}c {\hbox{\texttt{:}}} C{\hbox{\texttt{.}}}o{\char`\}} {\char`\{}$\Phi${\hbox{\texttt{.}}}o c a{\char`\}}}

\Scribtexttt{}\mbox{\hphantom{\Scribtexttt{x}}}\Scribtexttt{{\char`\{}c{\textquotesingle} {\hbox{\texttt{:}}} C{\hbox{\texttt{.}}}o{\char`\}} (f {\hbox{\texttt{:}}} C{\hbox{\texttt{.}}}$\Rightarrow$ c c{\textquotesingle}) $\rightarrow$}

\Scribtexttt{}\mbox{\hphantom{\Scribtexttt{x}}}\Scribtexttt{$\exists$ ($\lambda$ {\char`\{}c{\textquotesingle}{\textquotesingle} {\hbox{\texttt{:}}} C{\hbox{\texttt{.}}}o{\char`\}} $\rightarrow$ $\exists$ ($\lambda$ (g {\hbox{\texttt{:}}} C{\hbox{\texttt{.}}}$\Rightarrow$ c{\textquotesingle} c{\textquotesingle}{\textquotesingle}) $\rightarrow$}

\Scribtexttt{}\mbox{\hphantom{\Scribtexttt{x}}}\Scribtexttt{$\exists$ ($\lambda$ {\char`\{}a{\textquotesingle}{\textquotesingle} {\hbox{\texttt{:}}} A{\hbox{\texttt{.}}}o{\char`\}} $\rightarrow$ $\exists$ ($\lambda$ {\char`\{}h {\hbox{\texttt{:}}} A{\hbox{\texttt{.}}}$\Rightarrow$ a a{\textquotesingle}{\textquotesingle}{\char`\}} $\rightarrow$}

\Scribtexttt{}\mbox{\hphantom{\Scribtexttt{x}}}\Scribtexttt{$\exists$ ($\lambda$ {\char`\{}safe{\hbox{\texttt{.}}}$\Rightarrow$ g{\char`\}} $\rightarrow$ $\Phi${\hbox{\texttt{.}}}$\Rightarrow$ (C{\hbox{\texttt{.}}}compose g f) h)))))}\end{SingleColumn}\end{SVerbatim}

\noindent Erasing dependencies, implicit arguments, compile{-}time and redundant information,
the content can be extracted as a function in a programming language with less precise types:

\noindent \begin{SVerbatim}\begin{SingleColumn}\Scribtexttt{observe {\hbox{\texttt{:}}} (f {\hbox{\texttt{:}}} C{\hbox{\texttt{.}}}$\Rightarrow$) $\rightarrow$ (g {\hbox{\texttt{:}}} C{\hbox{\texttt{.}}}$\Rightarrow$)}\end{SingleColumn}\end{SVerbatim}

\noindent In lay words, \Scribtexttt{observe} takes the interrupted fragment of concrete computation
and shows how to complete it into one that is observable as an abstract computation.

Similarly, the computational content of completeness is a function that allows to
control the concrete computation as if it were the abstract computation.
The computational content of liveness is a function that advances
the concrete computation enough to advance the abstract computation.
All these functions and more form an API that allows arbitrary
implementations of arbitrary languages to be treated
as first{-}class objects, usable and \emph{composable} at runtime.

\SNextTitlePlain{Simulating or Performing Effects}\sectionNewpage

\Ssection{Simulating or Performing Effects}{Simulating or Performing Effects}\label{t:x28part_x22Simulatingx5forx5fPerformingx5fEffectsx22x29}

Traditional reflection protocols\Autobibref{~[\hyperref[t:x28autobib_x22Brian_Cantwell_SmithProcedural_Reflection_in_Programming_Languages1982x22x29]{\AutobibLink{7}}]}
offer interfaces where only state can be reified,
and effects always happen as ambient side{-}effects,
except sometimes for limited ad hoc ways to catch them.
By contrast, when extracting a protocol from a categorical specification,
it becomes obvious that effects too deserve first{-}class reification,
being the arrows of the reified computation category.

One simple way of reifying effects is as a journal recording I/O
that happened during the computation {---}
or would happen were the computation to actually run (or run again).
More abstract representations can be symbolic,
at a higher{-}level than a low{-}level logger could record;
they could be monadic functions or arbitrary Kleisli arrows.
In the end, there are two complementary approaches in which
effects are either \emph{simulated} or \emph{performed}.
Two functions \Scribtexttt{simulate} and \Scribtexttt{perform} may translate one approach into the other:
but while \Scribtexttt{perform} can be written on top of any expressive enough system (at a cost),
achieving \Scribtexttt{simulate} requires rewriting the entire system
if the existing implementation does not offer a suitable reflection protocol.

Now the reflection protocol itself includes effects beyond
those of the computations being reified and reflected {---}
if only partiality and its dual non{-}determinism.
Implementations and interpretations are partial and may fail on some nodes or arrows.
And even if a computation is itself deterministic or at least confluent,
running or advancing it includes non{-}determinism as to
how much work will be done according to what evaluation strategy.
A logical specification of the protocol must therefore expose these effects.

\SNextTitlePlain{Applications}\sectionNewpage

\Ssection{Applications}{Applications}\label{t:x28part_x22Applicationsx22x29}

The protocol, thanks to its crucial notion of observability, enables navigating up and down
a computation{'}s semantic tower \emph{while it is running}.
Developers can then zoom in and out of levels of abstraction
and focus their tools on the right level for whatever issue is at hand, neither too high nor too low.
Computations can be migrated from one underlying implementation to the other,
one machine or configuration to the other {---} changing a running engine.
Recovering an abstractly observable safe point also enables
safely killing threads and upgrading code,
thus achieving a robustness that only Erlang\Autobibref{~[\hyperref[t:x28autobib_x22Joe_Armstrong_and_Robert_Virding_and_Claes_Wikstrxf6m_and_Mike_WilliamsConcurrent_Programming_in_ERLANG1993x22x29]{\AutobibLink{8}}]} can currently provide.
Code instrumentations can be seen as the categorical opposites of Natural Transformations;
they can be written in a generic way, added to running code, configured independently from code;
they can provide orthogonal persistence, access control, time{-}travel debugging, and
other capabilities to all languages. etc.

Each of these applications has been done before, but in heroic ways,
available only to one implementation of one language,
using some ad hoc notion of safe points
(PCLSRing\Autobibref{~[\hyperref[t:x28autobib_x22Alan_BawdenPCLSRingx3a_Keeping_Process_State_Modular1989x22x29]{\AutobibLink{1}}]}, Garbage Collection\Autobibref{~[\hyperref[t:x28autobib_x22Paul_Rx2e_WilsonUniprocessor_Garbage_Collection_Techniques1992x22x29]{\AutobibLink{9}}]}, etc.).
The promise of this runtime reflection protocol is to achieve these applications
in comparatively simple yet general and \emph{composable} ways, and made available universally:
tools such as shells, debuggers, or code instrumentations,
can then work on all possible implementations of all languages,
specialized using e.g. typeclasses.

Finally, rooting a reflection protocol in formal methods means
it is now possible to reason about metaprograms, and maybe even feasably prove them correct;
they no longer need to invalidate semantic reasoning, nor to introduce unmanageable complexity.

\SNextTitlePlain{Conclusion and Future Work}\sectionNewpage

\Ssection{Conclusion and Future Work}{Conclusion and Future Work}\label{t:x28part_x22Conclusionx5fandx5fFuturex5fWorkx22x29}

The ideas above remain largely unimplemented.
But they already provide a new and promising way of looking at
either the semantics of implementations or the design of reflection protocols
{---} and more importantly, at the synergy between those two estranged fields.
My plan is to further implement the protocol in Gambit Scheme:
it already implements observability and migration at the level of its GVM\Autobibref{~[\hyperref[t:x28autobib_x22Marc_FeeleyCompiling_for_Multix2dlanguage_Task_Migration2015x22x29]{\AutobibLink{2}}]},
and there is a Racket{-}like module system called Gerbil to develop closed languages on top of it.

See my presentation at \href{https://youtu.be/heU8NyX5Hus}{https{\hbox{\texttt{:}}}//youtu{\hbox{\texttt{.}}}be/heU8NyX5Hus}.

\SNextTitlePlain{Bibliography}\sectionNewpage

\Ssectionstarx{Bibliography}{Bibliography}\label{t:x28part_x22docx2dbibliographyx22x29}

\begin{bigtabular}{@{\bigtableleftpad}l@{}l@{}}
\hbox{\Autocolbibnumber{[1]}} &
\hbox{\Autobibtarget{\label{t:x28autobib_x22Alan_BawdenPCLSRingx3a_Keeping_Process_State_Modular1989x22x29}\Autocolbibentry{Alan Bawden. PCLSRing: Keeping Process State Modular. 1989.}}} \\
\hbox{\Autocolbibnumber{[2]}} &
\hbox{\Autobibtarget{\label{t:x28autobib_x22Marc_FeeleyCompiling_for_Multix2dlanguage_Task_Migration2015x22x29}\Autocolbibentry{Marc Feeley. Compiling for Multi{-}language Task Migration. 2015.}}} \\
\hbox{\Autocolbibnumber{[3]}} &
\hbox{\Autobibtarget{\label{t:x28autobib_x22William_Ax2e_HowardThe_formulaex2dasx2dtypes_notion_of_construction1980x22x29}\Autocolbibentry{William A. Howard. The formulae{-}as{-}types notion of construction. 1980.}}} \\
\hbox{\Autocolbibnumber{[4]}} &
\hbox{\Autobibtarget{\label{t:x28autobib_x22Robin_MilnerAn_Algebraic_Definition_of_Simulation_between_Programs1971x22x29}\Autocolbibentry{Robin Milner. An Algebraic Definition of Simulation between Programs. 1971.}}} \\
\hbox{\Autocolbibnumber{[5]}} &
\hbox{\Autobibtarget{\label{t:x28autobib_x22Ulf_NorellDependently_typed_programming_in_Agda2008x22x29}\Autocolbibentry{Ulf Norell. Dependently typed programming in Agda. 2008.}}} \\
\hbox{\Autocolbibnumber{[6]}} &
\hbox{\Autobibtarget{\label{t:x28autobib_x22Franxe7oisx2dRenxe9_RideauReconciling_Semantics_and_Reflection2018x22x29}\Autocolbibentry{Fran\c{c}ois{-}Ren\'{e} Rideau. Reconciling Semantics and Reflection. 2018.}}} \\
\hbox{\Autocolbibnumber{[7]}} &
\hbox{\Autobibtarget{\label{t:x28autobib_x22Brian_Cantwell_SmithProcedural_Reflection_in_Programming_Languages1982x22x29}\Autocolbibentry{Brian Cantwell Smith. Procedural Reflection in Programming Languages. 1982.}}} \\
\hbox{\Autocolbibnumber{[8]}} &
\hbox{\Autobibtarget{\label{t:x28autobib_x22Joe_Armstrong_and_Robert_Virding_and_Claes_Wikstrxf6m_and_Mike_WilliamsConcurrent_Programming_in_ERLANG1993x22x29}\Autocolbibentry{Joe Armstrong and Robert Virding and Claes Wikstr\"{o}m and Mike Williams. Concurrent Programming in ERLANG. 1993.}}} \\
\hbox{\Autocolbibnumber{[9]}} &
\hbox{\Autobibtarget{\label{t:x28autobib_x22Paul_Rx2e_WilsonUniprocessor_Garbage_Collection_Techniques1992x22x29}\Autocolbibentry{Paul R. Wilson. Uniprocessor Garbage Collection Techniques. 1992.}}}\end{bigtabular}

\postDoc
\end{document}